\documentclass[a4paper,french]{rnti}
\usepackage{graphicx}
\titrecourt{}

\nomcourt{M. Sassi et S. Ben Bdira}
\titre{Traitement approximatif des requêtes flexibles avec groupement d'attributs et jointure}
\auteur{Minyar Sassi-Hidri\affil{1},
        Soukaina Ben Bdira\affil{2}\
}
\affiliation{
    Université Tunis El Manar\\Ecole Nationale d'Ingénieurs de Tunis\\BP. 37 Le Belvédère 1002 Tunis, Tunisia\\
          \{\affil{1}minyar.sassi,\affil{2}soukaina.benbdira\}@enit.rnu.tn\\

     }

\resume{
Cet article adresse le problème de traitement approximatif sur les requêtes
flexibles de la forme SELECT-FROM-WHERE-GROUP BY avec jointure
et propose un cadre flexible pour laggregation en ligne favorisant le temps
de réponse au détriment de lexactitude du résultat.
}
\summary{%
This paper addresses the problem of approximate processing for flexible queries in the form SELECT-FROM-WHERE-GROUP BY with join condition. It offers a flexible framework for online aggregation while promoting response time at the expense of result accuracy.
}
\begin{document}
\section{Introduction}
Les requêtes approximatives (RA) représentent une solution pertinente qui permet d'améliorer le temps de réponse aux dépens de l'exactitude. Celles-ci sont adaptées aux requêtes avec agrégats pour lesquelles la précision au dernier décimal n'est pas exigée. La contribution de ce travail est le calcul en ligne de l'agrégation pour les requêtes flexibles avec groupement d'attributs et jointure en se basant sur l'Analyse Formelle de Concepts (AFC) de \cite{will:1982ID5} et le formalisme des sous-ensembles flous.
\section{Agrégation en ligne des requêtes flexibles avec groupement d'attributs et jointure}
La première de notre approche consiste à générer la base de connaissances (BC) à partir de la base de données (BD). Elle est assurée par une procédure de classification non supervisée \cite{sassi12} sur les attributs relaxables\footnote{Attributs que les utilisateurs peuvent utiliser dans un prédicat de comparaison contenant un terme linguistique.}. La deuxième phase comporte deux étapes. La première est une réécriture de la requête qui aura la forme suivante d'une RA:

{\em \textbf{SELECT} Fonction(Attribut), DegréConfiance As Confidence,}

{\em FonctionInterval(DegréDeConfiance) \textbf{FROM} Tables}

{\em \textbf{WHERE} Attribut1 IS ConditionFlexible1 [And Attribut2 IS ConditionFlexible2...]} 

{\em \textbf{And} Table1.attribut=Table2.attribut... \textbf{GROUP BY} Table.Attribut;}

L'étape suivante consiste à construire un échantillon par jointure à partir de la BC. En effet, au lieu d'interroger toute la BC qui contient des milliers d'enregistrements, nous  interrogeons un échantillon de la BC qui est constitué par un sous ensemble de tuples de la BD, ce qui permet d'améliorer le temps de réponse. Cet échantillon sera transformé en un treillis de concepts, noyau de l'AFC, qui sera la clé du parcours et du calcul de la fonction d'agrégation ainsi que du taux d'erreur de cet échantillon. Pour le  calcul de l'agrégation, nous adoptons les fonctions d'agrégation définies initialement par \citet{hass99} et étendus par \citet{sassi12}.
\section{\'Evaluation}
\begin{figure}[h!]
\begin{center}
 \includegraphics[scale=0.2]{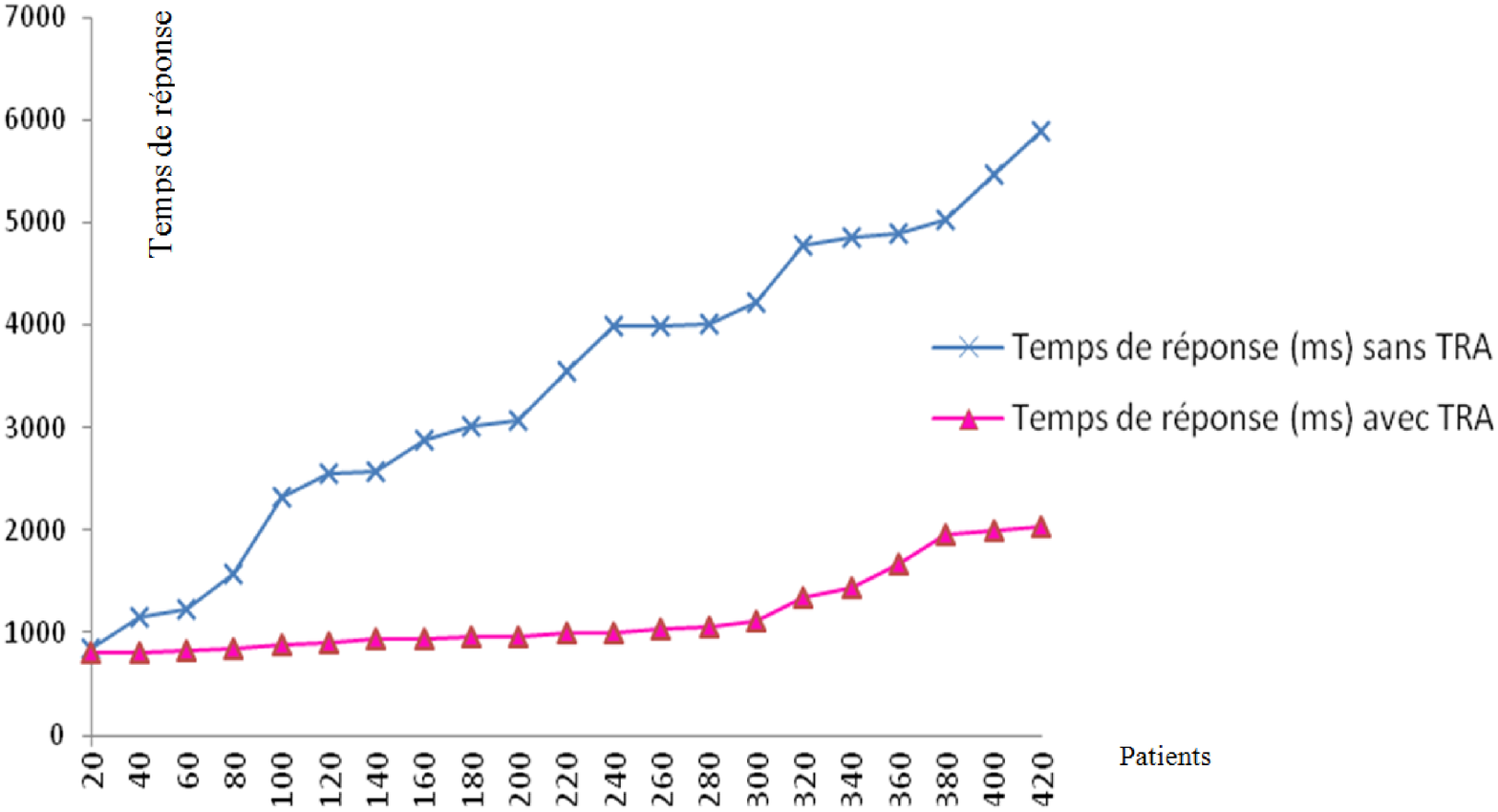}
 \caption{Comparaison entre l'interrogation flexible avec et sans traitement approximatif.} \label{fig1}
\end{center}
\end{figure}

Nous avons utilisé un exemple de  jeu de données qui gèrent un ensemble de patients afin d'étudier les facteurs de risque d'athérosclérose. Considérons la requête suivante "Lister le nombre de décès régulièrement alcooliques des patients scolaires par année". Les expérimentations faites prouvent que l'approche proposée favorise le temps de réponse au détriment de l'exactitude du résultat obtenu suite à une requête flexible. La figure \ref{fig1} présente la variation du nombre de patients par rapport au temps de réponse.
\vspace{-0.6cm}

Comme le montre la figure \ref{fig1}, le temps de réponse sans TRA est de l'ordre de 6 secondes (6000ms), tandis que, avec TRA, est de l'ordre de 2 secondes (2000ms). Pour tester l'exactitude de la réponse, nous extractions les valeurs exactes de la BD et nous les comparons avec celles obtenues par notre approche. Les résultats obtenus prouve l'efficience de l'approche proposée dans l'exactitude des réponses retournées.
\bibliographystyle{rnti}
\bibliography{biblio_exemple}
 %GATHER{c:/texmf_my/tex/latex/rnti/biblio_exemple.bib}
\Fr
\end{document}